\documentclass[ aps,
                prl,
                twocolumn,
                superscriptaddress
                ]{revtex4-1}
\usepackage{graphicx}
\usepackage{ulem}
\usepackage[dvipsnames]{xcolor}
\usepackage{amsmath}
\usepackage{amssymb}
\usepackage[colorlinks=true,urlcolor=blue,citecolor=blue,linkcolor=magenta]{hyperref}

\begin{document}
\title{Phase-Space Inequalities Beyond Negativities}

\author{Martin Bohmann}\email{martin.bohmann@oeaw.ac.at}
\affiliation{QSTAR, INO-CNR, and LENS, Largo Enrico Fermi 2, I-50125 Firenze, Italy}
\affiliation{Institute for Quantum Optics and Quantum Information - IQOQI Vienna, Austrian Academy of Sciences, Boltzmanngasse 3, 1090 Vienna, Austria}
\author{Elizabeth Agudelo}
\affiliation{Institute for Quantum Optics and Quantum Information - IQOQI Vienna, Austrian Academy of Sciences, Boltzmanngasse 3, 1090 Vienna, Austria}

\begin{abstract}
	We derive a family of inequalities involving different phase-space distributions of a quantum state which have to be fulfilled by any classical state.
	The violation of these inequalities is a clear signature of nonclassicality.
	Our approach combines the characterization of nonclassical effects via negativities in phase-space distributions with inequality conditions usually being formulated for moments of physical observables.
	Importantly, the obtained criteria certify nonclassicality even when the involved phase-space distributions are non-negative.
	Moreover, we show how these inequalities are related to correlation measurements.
	The strength of the derived conditions is demonstrated by different examples, including squeezed states, lossy single-photon states, and even coherent states.
\end{abstract}
\date{\today}
\maketitle

\paragraph{Introduction.---}	
    Our present capabilities in the preparation and control of quantum systems is revolutionizing our way of measuring, communicating, and computing. 
    These scientific and technological developments are based on the fact that quantum systems possess some properties that are impossible to describe with classical theories.
    Well-known examples for such genuine quantum features of the single-mode systems are photon antibunching \cite{Carmichael_1976,Kimble_1976,kimble_1977}, sub-Poissonian photon-number distributions \cite{mandel_1979,Zou_1990}, and squeezing \cite{Yuen_1976,Walls_1983,Caves_1985, Loudon_1987,Dodonov_2002}.
    As such nonclassical features are a resource for quantum technologies, and it is crucial to have efficient tools for their faithful detection and certification.

    One possibility of identifying genuine quantum features is using the framework of quasiprobability distributions.
    In this way, the incompatibility of the studied quantum system with classical physics is indicated through negativities in its quasiprobability distributions.
    This concept has a long-standing tradition in quantum optics where negativities in the Glauber-Sudarshan $P$ function \cite{sudarshan_1963,glauber_1963}, i.e., the state expansion in terms of a statistical mixture of (classical) coherent states, forms the very definition of nonclassicality \cite{titulaer_1965,mandel_1986}.
    Other types of widely used phase-space distributions such as the Wigner $W$ \cite{wigner_1932} and Husimi $Q$ \cite{husimi_1940} functions are related to the $P$ function through a convolution with a Gaussian kernel, forming the family of $s$-parametrized phase-space distributions \cite{cahill_1969a,cahill_1969b}.
    
    Quasiprobability distributions are also remarkably successful in other areas of modern quantum  physics.
    They are, for example, used in the context of entanglement \cite{sperling_2009,sperling_2019a}, contextuality \cite{spekkens_2008}, nonlocality \cite{lee_2009}, or general quantum coherence \cite{sperling_2018,sperling_2019}.
    Additionally, negativities in quasiprobability distributions indicate quantum resources for quantum information protocols \cite{veitch_2012,rahimi-keshari_2016,shahandeh_2017}.
    
    Alternative approaches for the detection and certification of genuine quantum characteristics are inequality conditions involving expectation values of different observables.
    These include conditions based on the Cauchy-Schwartz inequality \cite{agarwal_1988}, marginal distributions \cite{agarwal_1993}, uncertainty relations \cite{hillery_1987}, the matrix of moments approach  \cite{agarwal_1992,shchukin_2005a,shchukin_2005b,vogel_2008}, and others \cite{klyshko_1996,rivas_2009}.
    For a summary and comparison of different conditions, see Ref. \cite{miranowicz_2010}.
    On the one hand, such inequality conditions are in most cases easier to implement as only a few measurements are needed, in consequence, a full state reconstruction is not required.
    On the other hand, they only provide a nonclassicality test and are not sensitive towards all quantum features.
    
    In this Letter, we provide an unified approach that combines the certification of quantum correlations through quasiprobability distributions and inequality conditions.
    In particular, we introduce a family of nonclassicality inequality conditions for phase-space distribution functions.
    The present approach unifies the certification of nonclassicality through the state's phase-space distributions and inequality conditions typically involving expectation values of different observables.
    Our inequality criteria relate different phase-space representations of the state to each other, and can verify nonclassicality even if the involved phase-space distributions are smooth, non-negative functions, i.e., experimentally accessible ones.
    Furthermore, we demonstrate how our criteria can be implemented through correlation measurements that are widely used in quantum-optical experiments.
	
\paragraph{Phase-space distribution inequalities.---}
    We start with the diagonal quantum state representation through the Glauber-Sudarshan $P(\beta)$ distribution in the coherent state basis \cite{sudarshan_1963,glauber_1963},
	\begin{align}\label{eq:P}
		\hat \rho=\int d^2\beta\, P(\beta)|\beta\rangle\langle\beta|.
	\end{align}
	The $P$ function belongs to a family of phase space quasidistributions of the state that is typically characterized by a real parameter $s$ \cite{cahill_1969a}.
	In general, any $s$-parametrized phase-space function with $s{<}1$ can be expressed through the normal-ordered expectation value 
	\begin{align}\label{eq:Ps}
		P(\alpha;s)=\frac{2}{\pi(1-s)}\left\langle {:}\exp\left(-\frac{2}{1-s}\hat n(\alpha)\right){:}\right\rangle,
	\end{align}
	where $\hat n(\alpha)=\hat D(\alpha)\hat a^\dagger \hat a \hat D(\alpha)^\dagger$ is the displaced photon-number operator, $\hat D(\alpha)$ is the coherent displacement operator, and ${:}\dots{:}$ denotes the normal-order prescription; cf., e.g., Ref. \cite{vogel_2006}.
	Using Eq.~\eqref{eq:P}, we can write the expectation value in Eq.~\eqref{eq:Ps} in terms of the $P$ function,
	\begin{align}
		P(\alpha;s)=\frac{\eta_s}{\pi}\int d^2\gamma\, e^{(-k\eta_s|\gamma|^2)}e^{(-(1-k)\eta_s|\gamma|^2)} P(\gamma+\alpha),
	\end{align}
	where we introduced the parameter $\eta_s=2/(1-s)$, changed the integration variable ($\gamma=\beta-\alpha$), and split the exponential function into two functions parametrized by $k\in(0,1)$.
	Note that $\eta_s$ is always positive and, therefore, both exponential functions are decaying monotonically with $|\gamma|$.
	This monotonicity allows one to apply Chebyshev's integral inequality (see, e.g., Ref. \cite{mitrinovic_1970}), which yields
	\begin{align}
		P(\alpha;s)\stackrel{\mathrm{cl}}\geq & \, \,\frac{\eta_s}{\pi}\int d^2\gamma \exp\left(-k\eta_s|\gamma|^2\right) P(\gamma+\alpha)\nonumber\\
		&\times\int d^2\gamma \exp\left(-(1-k)\eta_s|\gamma|^2\right) P(\gamma+\alpha).
	\end{align}
	This condition is fulfilled for any quantum state that can be expressed in terms of a non-negative (classical) $P$ (for detailed steps on the derivation, we refer to the Appendix).
	Subsequently, this inequality can be  written in terms of different $s$-parametrized phase-space functions as
	\begin{align}\label{eq:inequalityP}
		P(\alpha;s)-\frac{\pi}{(1-k)k\eta_s} P(\alpha;s_k) P(\alpha;s_{(1-k)})\stackrel{\mathrm{cl}}\geq 0,
	\end{align}
	with $s_k=1-2/(k\eta_s)=1-(1-s)/k$.
	Interestingly, this result relates different $s$-parametrized phase-space distributions to each other and forms a nonclassicality test through its violation.
	For coherent states, which represent the boundary between the sets of classical and nonclassical states, the equality is attained.
	In addition, in order to certify nonclassicality, it is sufficient to find one combination of the parameters $s$ and $k$ for which this inequality is violated at least in one point $\alpha$ of the phase space.
	Note that the involved phase-space distributions are experimentally accessible, they can be pointwise sampled via unbalanced homodyne detection \cite{wallentowitz_1996} or reconstructed using other measurement strategies \cite{welsch_1999,lvovsky_2009}.
	
	To get a better understanding of the obtained result [Eq.  \eqref{eq:inequalityP}], we consider the case of $s=0$ and $k=1/2$, which yields
	\begin{align}\label{eq:WQ}
		W(\alpha)-2\pi\, Q(\alpha)^2\stackrel{\mathrm{cl}}\geq 0,
	\end{align}
	where $W(\alpha)=P(\alpha;0)$ and $Q(\alpha)=P(\alpha;-1)$ are the Wigner and Husimi $Q$ function, respectively.
	This expression is of particular importance as it relates the most widely used phase-space distributions to each other.
	The Wigner and Husimi $Q$ functions are reconstructed routinely in quantum optics laboratories for various types of quantum states  \cite{smithey_1993,leibfried_1996,kanem_2005,deleglise_2008,vlastakis_2013,jeong_2014,strobel_2014,sychev_2017,hacker_2019,landon-cardinal_2018}.
	In particular, Eq.~\eqref{eq:WQ} sets a lower bound for the Wigner function of any classical state in terms of the square of the corresponding $Q$ function at any point in phase space.
	Therefore, it tells us how much classical phase-space distributions can change their values by changing the $s$ parameter (this aspect will be discussed in more detail below).
	Furthermore, we see that the classicality condition in Eq.~\eqref{eq:WQ} is always violated when $W(\alpha)<0$.
	This is not surprising as the negativities in the Wigner function by themselves are already sufficient signatures of nonclassicality.
	
	The inequality is, however, getting more interesting when we consider quantum states that are nonclassical but possess non-negative Wigner functions, such as squeezed states, which are nonclassical Gaussian states.
	In these cases, this inequality provides the possibility to certify nonclassicality despite the fact that all involved phase-space distributions are non-negative.

        \begin{figure}[b]
		\centering
		\includegraphics[width=0.45\columnwidth]{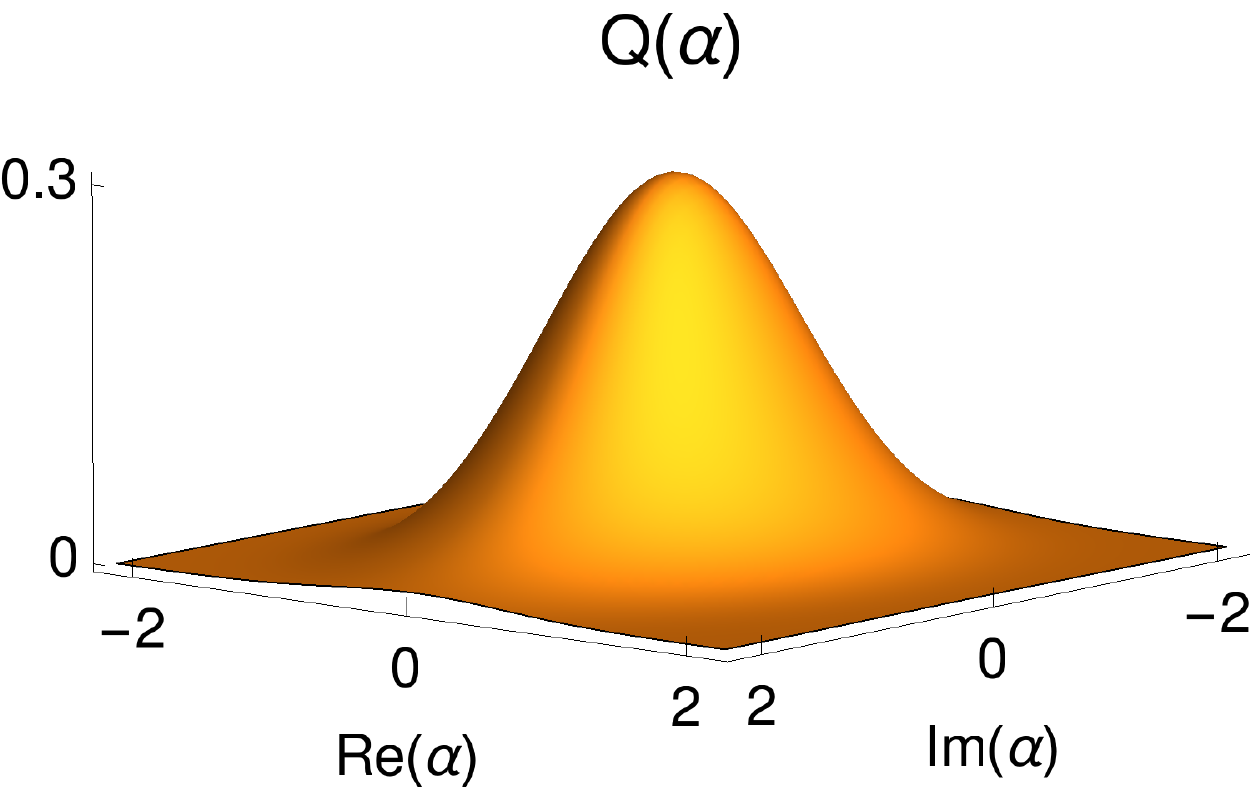}
		\includegraphics[width=0.45\columnwidth]{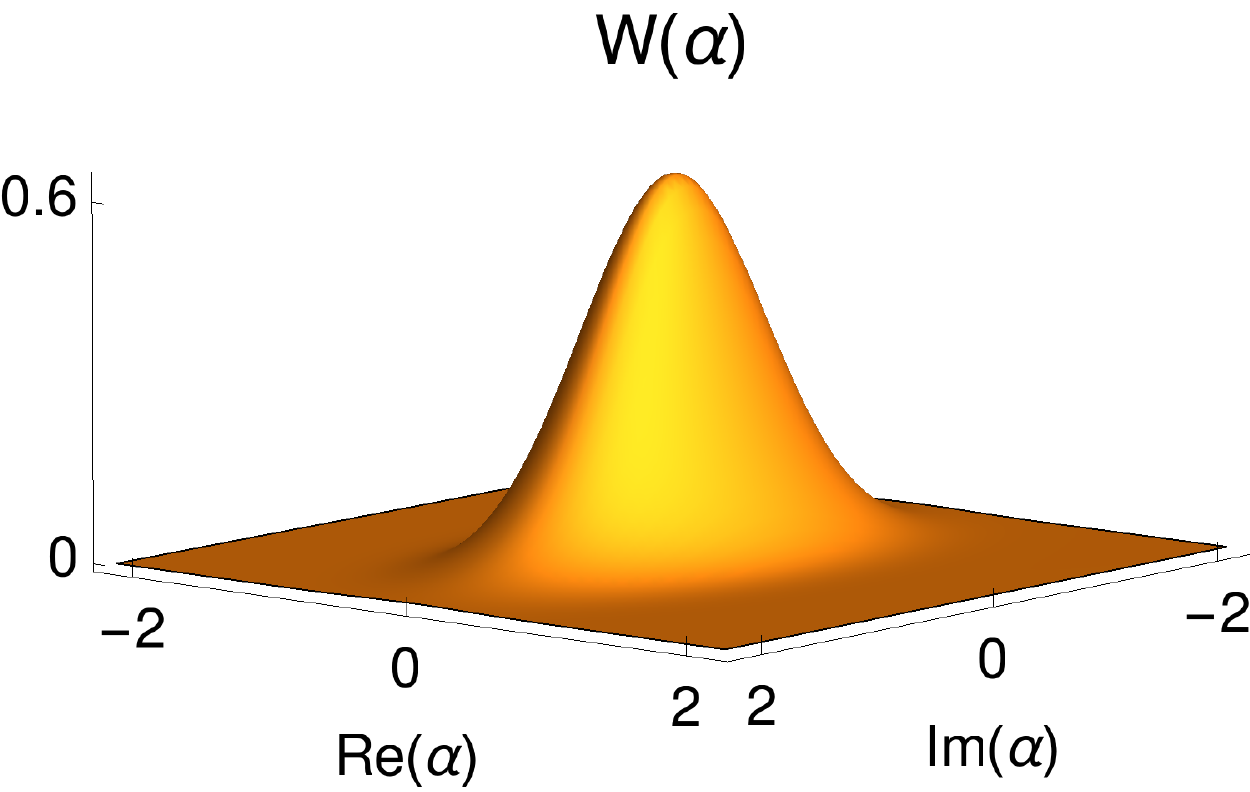}\\[2mm]
		\includegraphics[width=0.7\columnwidth]{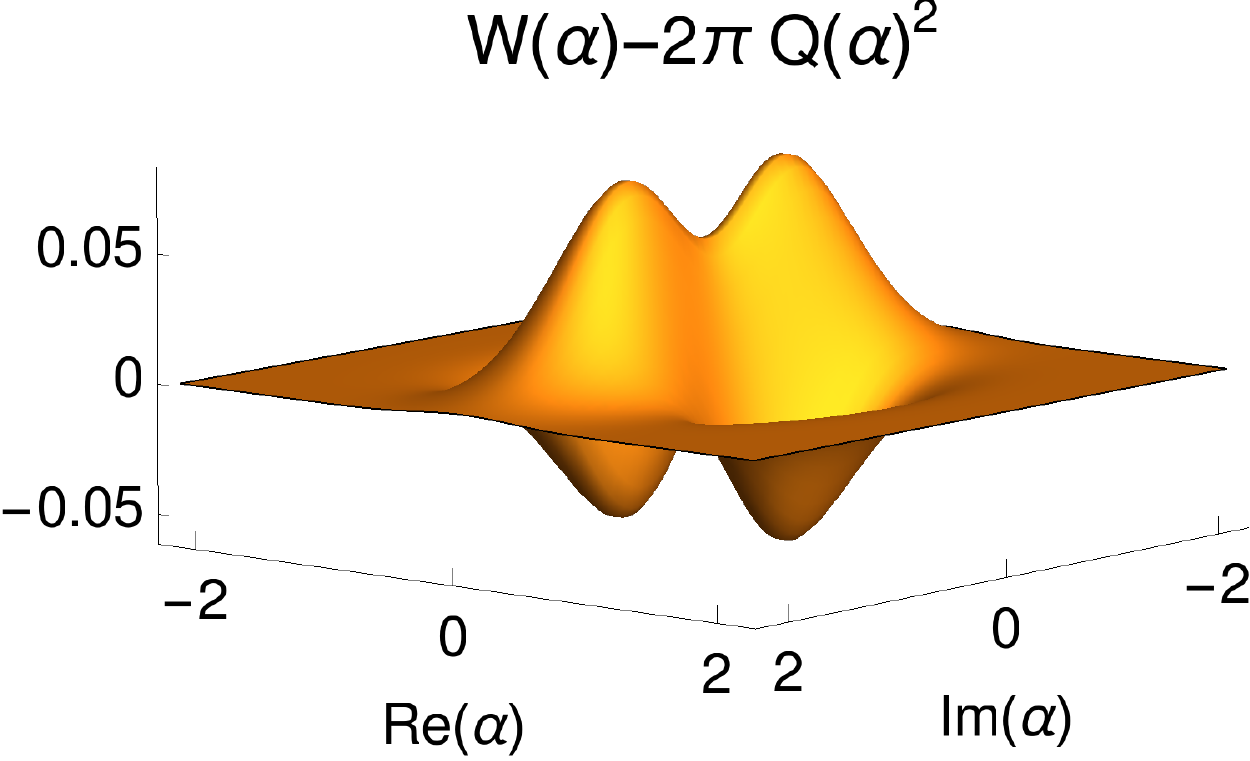}\caption{
		        The Wigner and $Q$ function of a squeezed state are shown (top) together with the left-hand side of Eq.~\eqref{eq:WQ} (bottom).
		        Negativities in the latter certify the nonclassicality of the state.
			}
	    \label{fig:QW}
    \end{figure}
    
\paragraph{Certifying nonclassical states.---}
    First,~let~us consider a single-mode squeezed vacuum state, which is defined as $|\xi\rangle=(\cosh{r})^{-1/2}$ $\sum_{n=0}^\infty (-e^{\varphi_\xi}\tanh{r})^n$ $\sqrt{(2n)!}/(2^n n!)|2n\rangle$ with $r=|\xi|$ and $\varphi_\xi=\arg(\xi)$.
    Squeezed states are known to have non-negative Gaussian Wigner functions; see, e.g., Ref. \cite{weedbrook_2012}.
    However, they are nonclassical states which allow for quantum enhanced applications, e.g., in quantum metrology \cite{grote_2013} or quantum information \cite{weedbrook_2012}.
    In Fig.~\ref{fig:QW}, the Wigner function, the $Q$ function, and the left-hand side (lhs) of inequality \eqref{eq:WQ} are plotted for a squeezed state with $\xi=0.3$, corresponding to $2.7$dB of squeezing.
    Both the Wigner and $Q$ function of the squeezed state are non-negative, and we cannot directly infer nonclassicality of the state from these phase-space distributions.
    In contrast, when one evaluates the lhs of Eq.~\eqref{eq:WQ}, one clearly sees negative values in Fig.~\ref{fig:QW}.
    These negativities imply a violation of the inequality \eqref{eq:WQ} certifying the nonclassicality of the squeezed state.
    
\begin{figure}[t]
		\centering
		\includegraphics[width=0.8\columnwidth]{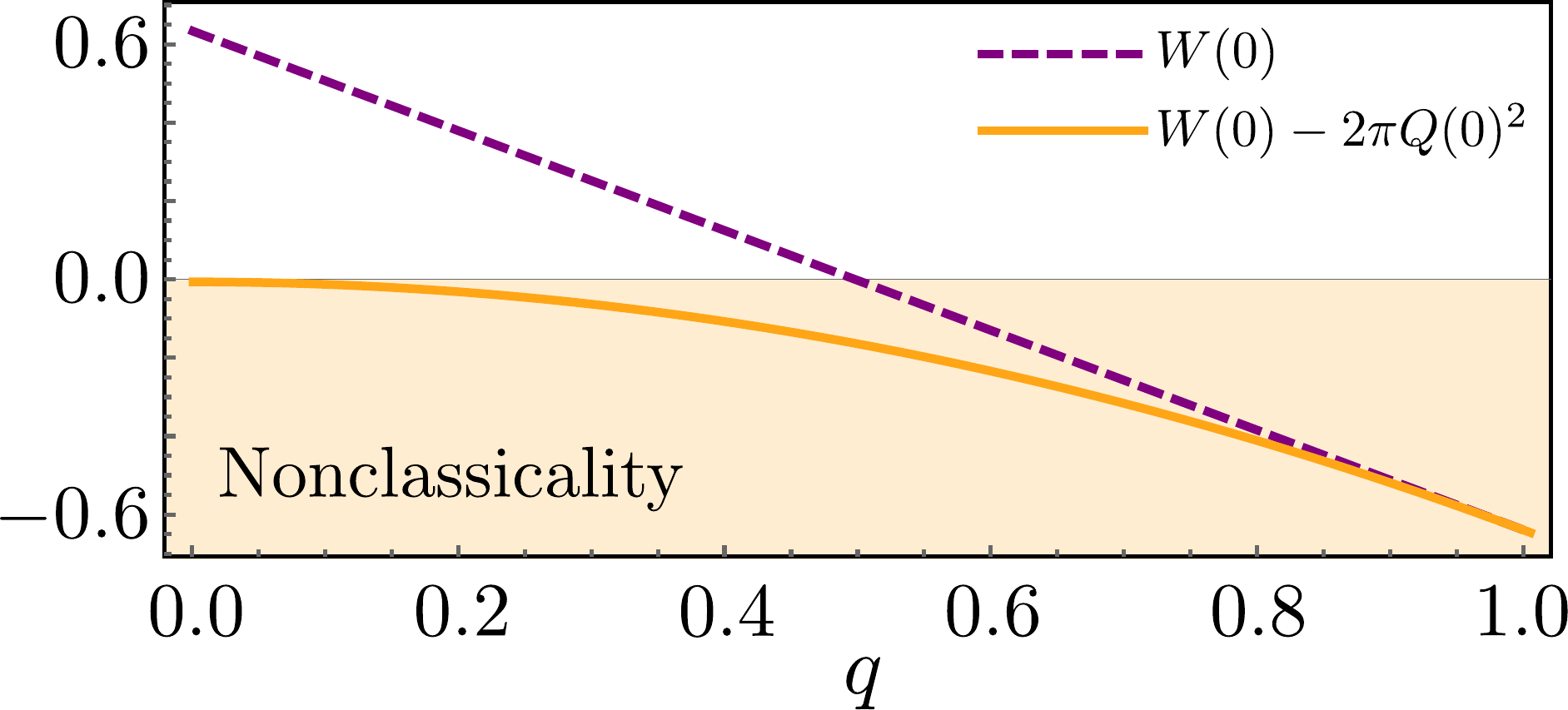}
			\caption{
			    The phase-space relation in Eq.~\eqref{eq:WQ} (solid line) and the Wigner function (dashed line) of a lossy photon state are plotted at the origin of phase space ($\alpha=0$) as functions of the loss parameter $q$.
				Negative values indicate nonclassicality.
			}
		\label{fig:lossyPhoton}published version, now includes supplemental material 
	\end{figure}

    As a second example, we consider a lossy photon state---a non-Gaussian state---, $\hat\rho_q=q|1\rangle\langle 1|+(1-q)|0\rangle\langle 0|$, which is parametrized through the loss parameter $q\in [0,1]$.
    For $q=1$, $\hat \rho_q$ is a pure single-photon state and with decreasing $q$ the loss influence is increasing.
    For the lossy single-photon state with $q>0.5$, it is known that its Wigner function has a negativity at the origin of the phase space \cite{lvovsky_2001}, which directly certifies its nonclassicality.
    In Fig.~\ref{fig:lossyPhoton}, we compare the value of the Wigner function with the one of the lhs of the condition \eqref{eq:WQ} at $\alpha=0$.
    The here derived phase-space inequality condition is capable of identifying the nonclassicality of $\hat \rho_q$ for any nontrivial $q$ ($q\neq 0$).
    Hence, it verifies nonclassicality even in the case of strong losses or low detection efficiencies which makes it a robust and effective tool for many experimental scenarios.
    
\begin{figure}[t]
		\centering
		\includegraphics[width=0.75\columnwidth]{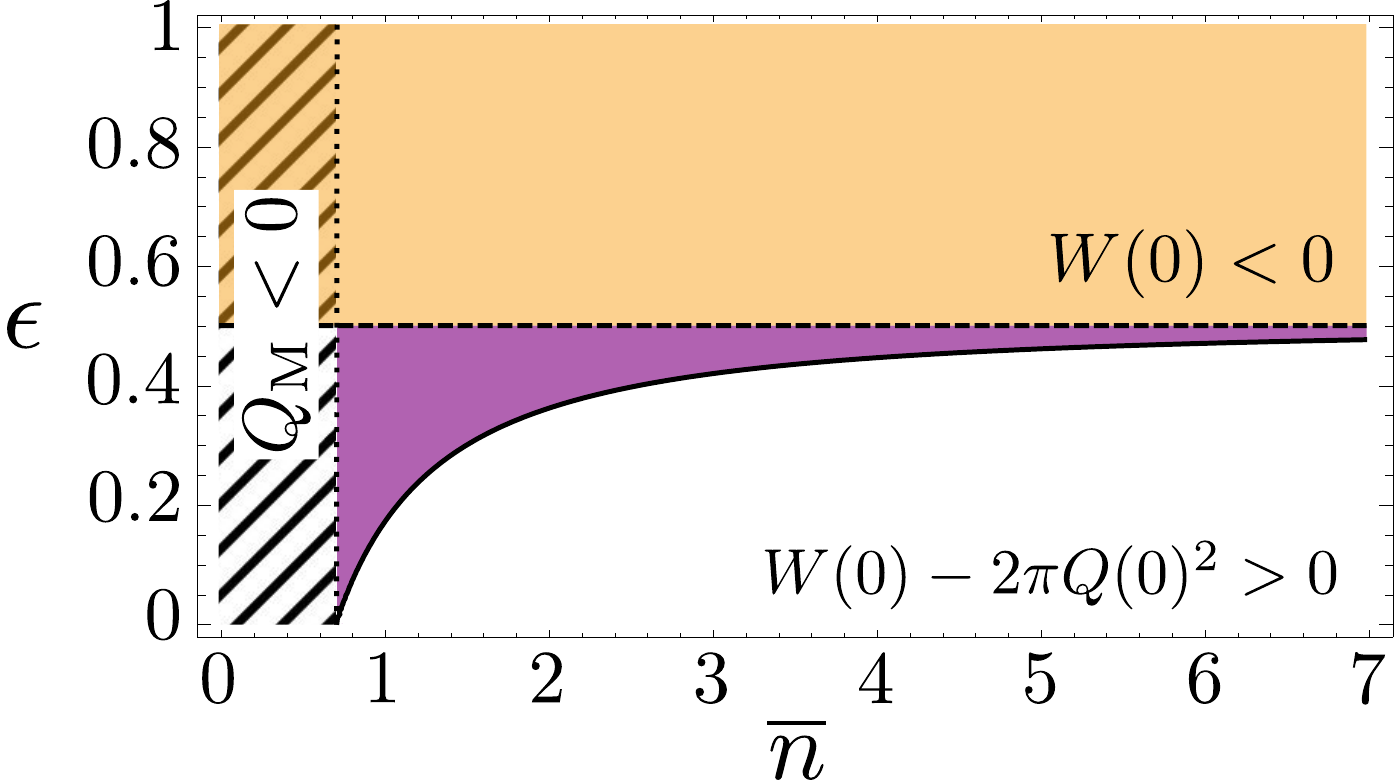}
			\caption{Nonclassicality certification of the lossy SPATS in terms of losses ($1-\epsilon$) and thermal photon number ($\overline{n}$).
			Nonclassicality is detected via negativities of the Mandel $Q_\mathrm{M}$ parameter for $\overline{n}<0.707$ (dotted line) and via negative Wigner function for $\epsilon>0.5$ (dashed line).
			Above the solid curve, inequality Eq. \eqref{eq:WQ} is violated.
			}
		\label{fig:lossySPATS}
	\end{figure}

	Finally, we compare the derived criteria with well-established nonclassicality tests.
	For this purpose, we consider the single-photon added thermal state (SPATS), $\mathcal{N}\hat a^\dagger\hat\rho_{\mathrm{th}}\hat a$, with $\hat\rho_{\mathrm{th}}=1/(\overline{n}+1)\sum_{k=0}^\infty \left[\overline{n}/(\overline{n}+1)\right]^k |k\rangle\langle k|$ and $\mathcal{N}$ being the normalization constant \cite{agarwal_1992}.
	SPATSs are nonclassical states that are experimentally realizable \cite{zavatta_2007}.
	They do not exhibit quadrature squeezing, have a positive Mandel $Q_{\mathrm{M}}$ parameter \cite{mandel_1979} for $\overline{n}>0.707$ \cite{agarwal_1992}, and possess a positive Wigner function for losses above $50\%$ ($\epsilon<0.5$) \cite{kuehn_2014}.
	Furthermore, their characteristic function (Fourier transform of the $P$ function) does not violate the nonclassicality condition based on the Bochner theorem \cite{vogel_2000} for $\overline{n}>0.386$ \cite{kuehn_2014}.
	In Fig. \ref{fig:lossySPATS}, we show that condition Eq. \eqref{eq:WQ} can detect nonclassicality of the lossy SPATS in a wider parameter range (purple region) than all these well-established tests.
	This example shows that the introduced approach can be advantageous over established nonclassicality criteria.
	Note, however, that there are other ways to detect the nonclassicality of this state; cf., e.g., Refs. \cite{yadin_2016,debievre_2019,luo_2019}.
	
\begin{figure}[b]
		\centering
		\includegraphics[width=0.6\columnwidth]{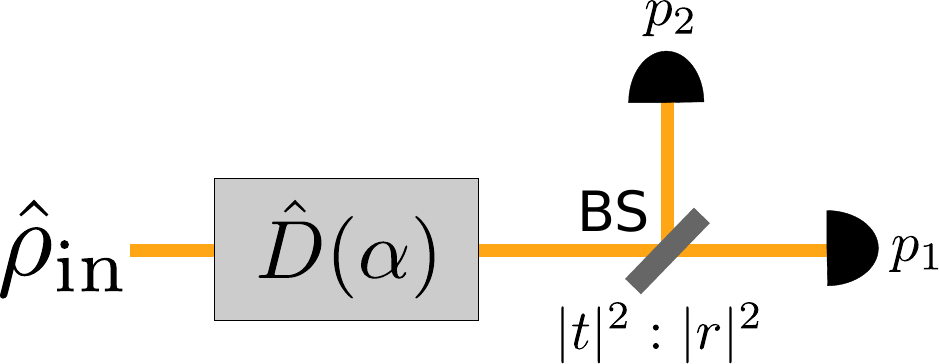}
			\caption{Schematics of the correlation measurement technique.
			}
		\label{fig:MPCat}
	\end{figure}

\paragraph{Relation to correlation measurements.--}
    Our phase-space-distribution inequalities are closely related to correlation measurements as depicted in Fig.~\ref{fig:MPCat}.
    For this purpose, we first recall that the $s$-parametrized phase-space functions \eqref{eq:Ps} can be directly sampled from the zero-count probability by unbalanced homodyne detection \cite{wallentowitz_1996} as 
	\begin{align}\label{eq:Pp}
		P(\alpha;s)=\frac{2}{\pi(1-s)}\,p(\alpha,\eta_s),
	\end{align}
	where $p(\alpha,\eta_s)$ is the probability to register zero photons \cite{kelley64,vogel_2006} given by  $p(\alpha,\eta_s)=\langle{:}\exp[-\eta_s \hat n(\alpha)]{:}\rangle$, and $\eta_s$ is the detector efficiency.
	Physical values of $\eta_s$ restrict which phase-space distributions are measurable in this manner.
	Therefore, measuring the zero-count probability can at best ($\eta_{s=-1}=1$) allow for the sampling of the $Q$ function.
	In realistic settings with quantum efficiencies less than 1, only phase-space functions which are even smoother than the $Q$ function, i.e., $P(\alpha;s{<}-1)$, are accessible from the zero-count probability.
	The question arises if we can still certify nonclassicality from such non-negative, smooth distributions.
	
	With Eq.~\eqref{eq:Pp} at hand, we show how the correlation measurement depicted in Fig.~\ref{fig:MPCat} can be used to directly sample our phase-space distribution inequalities.
    The input state $(\hat \rho_{\mathrm{in}})$ is displaced $[\hat D(\alpha)]$, split at a beam splitter (BS), and then at both outputs the zero-count probabilities are recorded.
    Taking into account the intensity splitting ratio of the beam splitter ($|t|^2{+}|r|^2=1$), the zero-count probability of the two detectors are given by $p_1(\alpha,\eta |r|^2)$ and $p_2(\alpha,\eta |t|^2)$ where $\eta$ is their detection efficiency.
	Furthermore, we can also consider $p_{1,2}(\alpha,\eta)$, which is the coincident zero-count probability.
	From this consideration, we see that the zero-count covariance,
	\begin{align}
	    p_{1,2}(\alpha,\eta)-p_1(\alpha,\eta |r|^2) p_2(\alpha,\eta |t|^2),
	\end{align}
	is in fact nothing else than the lhs of our phase-space distribution inequality \eqref{eq:inequalityP} multiplied by $\pi(1-s)/2$, with $s=1-2/\eta$ and $k=|t|^2$.
    Therefore, such a simple correlation measurement based on the zero-count probabilities of two detectors provides an easily accessible experimental test of Eq.~\eqref{eq:inequalityP}.
    Note also that such correlation conditions are not altered by incoherent noise, such as dark counts or stray light \cite{lipfert_2015,bohmann_2019}, which makes them robust against experimental noise. 
    One more question arises: is it possible to certify nonclassicality through such correlation measurements as the related phase-space distributions have always more Gaussian noise added than the $Q$ function?
    
\begin{figure}[b]
		\centering
		\includegraphics[width=0.7\columnwidth]{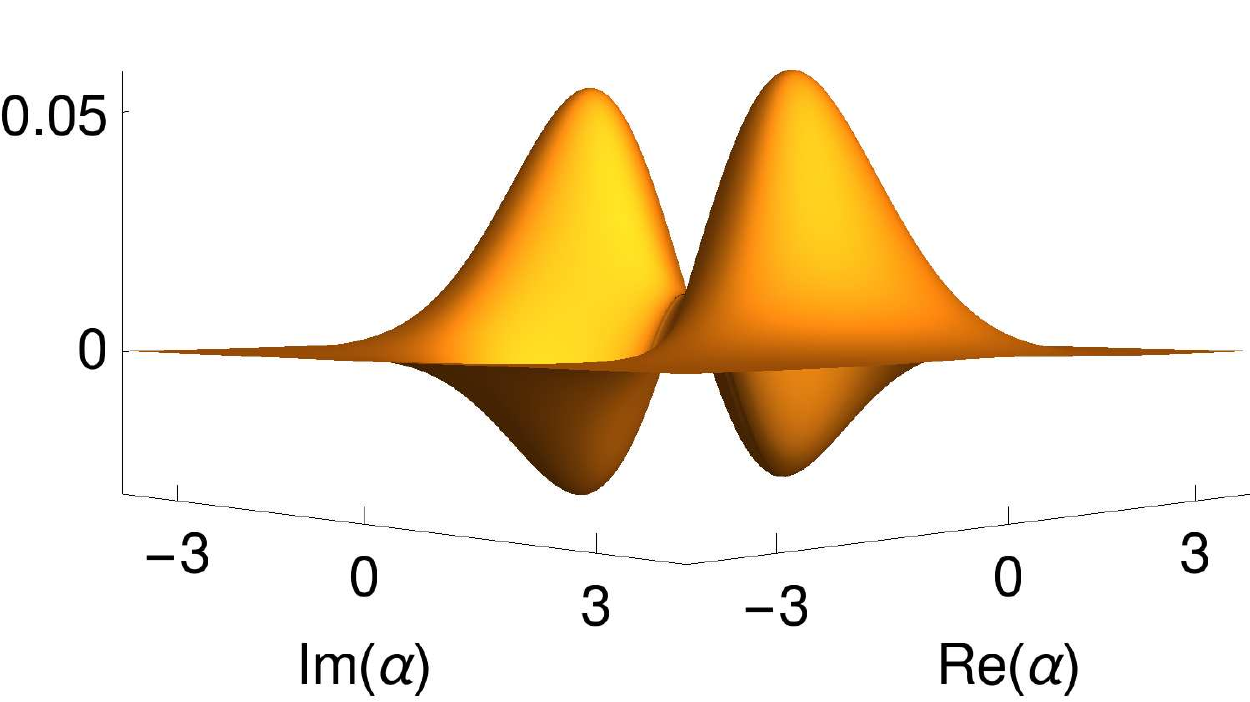}
			\caption{
			    Left-hand side of inequality \eqref{eq:inequalityP} ($s=-3$ and $k=1/2$) for a cat state with coherent amplitude $\omega=0.7$ sampled from the correlation measurement depicted in Fig.~\ref{fig:MPCat}.
			}
		\label{fig:MPCatb}
	\end{figure}

    We address this question by studying the example of a even coherent state, $\mathcal{N}(|\omega\rangle{+}|{-}\omega\rangle)$, where $|\pm \omega\rangle$ are coherent states and $\mathcal{N}=\{2[1{+}\exp(-2|\omega|^2)]\}^{1/2}$, which is symmetrically split $(|t|^2=1/2)$ and recorded with a quantum efficiency of $\eta=0.5$.
    This means that we can sample $s$-parametrized phase-space distribution with $s={-}7$ from the $p_1$ and $p_2$, and one with $s={-}3$ from the joint zero-count probability $p_{1,2}$.
    Using our inequality condition \eqref{eq:inequalityP}, we clearly certify nonclassicality in terms of negative values violating the classicality condition; as seen in Fig.~\ref{fig:MPCatb}.
    Therefore, we could show that the introduced family of phase-space distribution inequalities is capable of verifying nonclassicality, for some nonclassical states, even if the involved distributions are non-negative and very smooth $(s<-1)$.
    Simple zero-count correlation measurements are a feasible way for their implementation in experiments.
    
    This consideration can be extended to general multiplexing schemes \cite{paul_1996,kok_2001,achilles_2003,fitch_2003,castelletto_2007,schettini_2007,sperling_2015,bohmann_2018}, i.e., to the case in which the input state is split into an $N$-mode output state ($N\geq 2$).
    By applying Chebyshev's integral inequality \cite{mitrinovic_1970} $N{-}1$ times, one obtains the multimode generalization of Eq.~\eqref{eq:inequalityP}
	\begin{align}\label{eq:inequalityPgeneralization}
		P(\alpha;s)-\left(\frac{\pi}{\eta_s}\right)^{N-1}\prod_{i=1}^N k_i^{-1}  P(\alpha;s_{k_i}) \stackrel{\mathrm{cl}}\geq 0,
	\end{align}
	with $\sum_{i=1}^N k_i=1$ and $s=1-2/\eta$, which can be directly sampled from the zero-count probabilities of the different detection channels in a multiplexed detection scenario.
	Details on this generalization are provided in the Appendix.
	This connects the introduced phase-space distribution inequality approach to general multiplexing detection schemes---an experimental technique widely used in quantum optics and technology---and allows for the verification of nonclassical light in terms of the recorded zero-count events.

\paragraph{Discussion.---}
    The derivation of the phase-space distribution inequalities Eqs. \eqref{eq:inequalityP} and \eqref{eq:inequalityPgeneralization} rely on first principles only, assuring the universality of the whole approach.
    Importantly, these inequalities apply to any bosonic quantum system and are not limited to radiation fields.
    The violation of the inequalities provides a direct certification of nonclassicality.
    Furthermore, they connect two different approaches of certifying nonclassicality---through negativities in phase-space distributions and through the violation of inequalities for different observables---which results in an interesting and versatile tool for the characterization of quantum systems.
    In this context, it is important to stress that this approach is capable of revealing nonclassicality even if the involved phase-space distributions are all non-negative, smooth functions $(s<-1)$.
    Hence, one can avoid dealing with potentially ill-behaved distributions, e.g., featuring rapid oscillations or singularities (as can be typically found in quasiprobabilities with $s>0$), while still being able to verify nonclassicality.
    Furthermore, we remark again that it is sufficient to find one point in phase space for which the inequality is violated in order to certify nonclassicality.
    Consequently, it is not necessary to sample or reconstruct the involved distributions in the whole phase space to detect the nonclassical character of the studied system.
    The points above illustrate the practicability of the introduced approach.
 
    Equation \eqref{eq:inequalityP} provides a classical bound on how much the value of the $s$-parametrized distributions can vary with respect to changes in the $s$ parameter.
    For example, in the special case in Eq.~\eqref{eq:WQ}, the Wigner function of any classical state is bounded from below by the square of the $Q$ function (times a constant).
    Therefore, the derived inequalities are related to the recently introduced operator ordering sensitivity \cite{debievre_2019} which captures the sensitivity of $s$-parametrized phase-space distributions with respect to infinitesimal changes in $s$.
    
    Finally, we want to comment on some possible extension of the introduced approach.
    The here obtained inequalities can be extended to generalized phase-space distributions which are obtained via the convolution of the $P$ function with a general (non-Gaussian) kernel (cf., e.g., Ref. \cite{sperling_2019} for an overview) which, for example, includes filtered phase-space distributions \cite{kiesel_2010,agudelo_2013}.
    An extension to multimode scenarios and time-dependent systems \cite{agarwal_1970b,vogel_2008,krumm_2017} is also conceivable.
    Furthermore, we note that the derivation of our conditions relies on the prerequisite of the non-negativity of the distribution function.
    Therefore, it might be possible to adapt the presented approach, i.e., the application of Chebyshev's integral inequality, to other kinds of quantumness for which ``classical'' reference states form a convex set.
    This relates to the resource theory for quantum coherence \cite{streltsov_2017}, which is relevant for quantum technologies, and to its corresponding quasiprobability representation \cite{sperling_2018}.

\paragraph{Conclusions.---}
    We introduced inequality conditions involving different $s$-parametrized phase-space functions of a bosonic system whose violation is a direct indication of nonclassicality of its quantum state.
    This approach unites the verification of nonclassicality in terms of phase-space distributions and inequality conditions---the two prevalent ways for revealing nonclassicality of quantum states.
    The derivation of the family of inequality conditions is based on the Chebyshev's integral inequality and is directly connected to the definition of nonclassicality.
    We demonstrated the usefulness of the nonclassicality conditions by means of several examples, and we showed that even under poor circumstances, such as strong losses, nonclassicality can still be certified.
    We explicitly showed that the inequality conditions can be easily applied to and sampled from multiplexed correlation measurements, which reveals a connection between correlation measurements and phase-space distributions.
    The presented approach is, however, not limited to such measurements as it applies to any scenario in which phase-space distributions can be reconstructed or sampled.
    This assures a wide applicability in may experiments and studies in quantum science.
	
    The authors thank Jan Sperling for stimulating discussions and Werner Vogel for helpful comments.
    M.B. acknowledges financial support by the Leopoldina Fellowship Programme of the German National Academy of Science (LPDS 2019-01).
    E.A. acknowledges funding from the European Union's Horizon 2020 research and innovation programme under the Marie Sk\l{}odowska-Curie IF (InDiQE - EU project 845486).

\appendix
\begin{widetext}
\section{Appendix}
	This supplemental material is organized as follows. 
	In Sec.~\ref{sec:derivation}, we present the derivation of the phase-space distribution inequalities.
	A generalization of the correlation measurement scheme to multi-mode scenarios is provided in Sec.~\ref{sec:generalization}.

\section{Derivation of the inequality conditions}
\label{sec:derivation}
\subsection{Chebyshev’s integral inequality}
    For the derivation of our phase-space-distribution inequalities we will make use of the Chebyshev's integral inequality; see, e.g.,~\cite{mitrinovic_1970}.
    Therefore, we will briefly state and explain this inequality.
	Consider two functions $f$ and $g$ which are integrable and monotone in the same sense on $(a,b)$, and a positive function $p$ which is integrable on the same interval. 
	Then the Chebyshev's integral inequality
	\begin{align}
		\int_a^b p(x) f(x) g(x)dx \int_a^b p(x) dx \geq
		\int_a^b p(x) f(x)dx  \int_a^b p(x) g(x)dx,
	\end{align}
	holds.
	In the case that $p(x)$ is a probability distribution on $(a,b)$ the inequality reduces to 
	\begin{align}\label{eq:ChebyshevSupp}
		\int_a^b p(x) f(x) g(x)dx \geq
		\int_a^b p(x) f(x)dx  \int_a^b p(x) g(x)dx.
	\end{align}
	For deriving our nonclassicality conditions, $p(x)$ will be the displaced and phase-averaged $P$ function of a classical quantum state and $f$, $g$ are the normal-ordered expectation values which are related with the $s$-parametrized phase-space distributions.
	
\subsection{Derivation}
     Here, we present the detailed derivation of our phase-space-distribution inequality.
     We start from the expression of the $s$-parametrized phase-space functions \cite{cahill_1969a}
	\begin{align}\label{eq:sSupp}
		P(\alpha;s)=\frac{2}{\pi(1-s)}\left\langle {:}\exp\left(-\frac{2}{1-s}\hat n(\alpha)\right){:}\right\rangle
	\end{align}
	where $\hat n(\alpha)=\hat D(\alpha)\hat a^\dagger \hat a \hat D(\alpha)^\dagger=(\hat a^\dagger-\alpha^*)(\hat a-\alpha)$ is the displaced photon-number operator, $\hat D(\alpha)$ is the coherent displacement operator, and ${:}\dots{:}$ denotes the normal-order prescription; cf., e.g., \cite{vogel_2006}.
	We explicitly write this expectation value of in terms of the Glauber-Sudarshan $P$ function of the input state
	\begin{align}
		P(\alpha;s)=\frac{2}{\pi(1-s)}\int d^2\beta P(\beta) \exp\left(-\frac{2}{1-s} |\beta-\alpha|^2\right).
	\end{align}
	In a first step, we substitute $\gamma=\beta-\alpha$, introduce $\eta_s=2/(1-s)$, and split the exponential function of the integrand into two separate functions
	\begin{align}
		P(\alpha;s)=\frac{\eta_s}{\pi}\int d^2\gamma P(\gamma+\alpha) \exp\left(-k\eta_s |\gamma|^2\right) \exp\left(-(1-k)\eta_s |\gamma|^2\right),
	\end{align}
    with $k\in(0,1)$.
    We note that the exponential functions in integrand are not dependent on the phase but only on the amplitude of $\gamma$.
    Therefore, we can change to polar coordinates, $\gamma=r\exp(i\varphi)$ and $\alpha=r_\alpha\exp(i\varphi_\alpha)$, and rearrange the integral to 
    \begin{align}
		P(\alpha;s)&=\frac{\eta_s}{\pi}\int_0^\infty dr \tilde P(r;\alpha) \exp\left(-k\eta_s r^2\right) \exp\left(-(1-k)\eta_s r^2\right),\\
		\mathrm{with}\quad\tilde P(r;\alpha)&=\int_0^{2\pi} d \varphi r  P\left(\sqrt{r^2+r_\alpha^2+2r r_\alpha\cos (\varphi_\alpha-\varphi)},\varphi+\mathrm{arctan2}\left(r_\alpha\sin(\varphi_\alpha-\varphi),r+r_\alpha\cos(\varphi_\alpha-\varphi)\right)\right)
	\end{align}
	where the function $\mathrm{arctan2}$ is defined as 
	\begin{align}
	    \mathrm{arctan2}(y,x)=
	    \begin{cases} 
              \mathrm{arctan}(\frac{y}{x}) & \mathrm{if}\quad x>0\\
              \mathrm{arctan}(\frac{y}{x})+\pi & \mathrm{if}\quad x<0 \quad\mathrm{ and }\quad y\geq 0\\
              \mathrm{arctan}(\frac{y}{x}) -\pi & \mathrm{if}\quad x<0 \quad\mathrm{ and }\quad y< 0\\
              +\frac{\pi}{2}  & \mathrm{if}\quad x=0 \quad\mathrm{ and }\quad y>0\\
              -\frac{\pi}{2}  & \mathrm{if}\quad x=0 \quad\mathrm{ and }\quad y<0\\
              \mathrm{undefined} & \mathrm{if}\quad x=0\quad \mathrm{ and }\quad y=0        
              \end{cases}.
	\end{align}
	Here, it is important to stress that for any classical state $\tilde P(r;\alpha)$ is a probability distribution with respect to the variable $r$, i.e., $\tilde P(r;\alpha)\geq 0$  $\forall r$ and $\int_0^\infty dr \tilde P(r;\alpha)=1$.
	Now we have everything at hand in order to apply Chebyshev’s integral inequality \eqref{eq:ChebyshevSupp} to the above integral expression.
	In particular, by comparing the two equations, we identify $r$ with $k$, $\tilde P(r;\alpha)$ with $p(x)$, $\exp\left(-k\eta_s r^2\right)$ with $f(x)$, and $\exp\left(-(1-k)\eta_s r^2\right)$ with $g(x)$.
	Applying Eq.~\eqref{eq:ChebyshevSupp} then yields 
	\begin{align}
		P(\alpha;s)=&\frac{\eta_s}{\pi}\int_0^\infty dr \tilde P(r;\alpha) \exp\left(-k\eta_s r^2\right) \exp\left(-(1-k)\eta_s r^2\right)\\ &\stackrel{\mathrm{cl}}\geq
		\frac{\eta_s}{\pi}\int_0^\infty dr \tilde P(r;\alpha) \exp\left(-k\eta_s r^2\right)
		\int_0^\infty dr \tilde P(r;\alpha) \exp\left(-(1-k)\eta_s r^2\right),
	\end{align}
	which has to be fulfilled for any classical state, i.e., states with a non-negative $P$ function.
	By resubstituting and rearranging the above expression we obtain 
	\begin{align}
		P(\alpha;s)\stackrel{\mathrm{cl}}\geq\frac{\eta_s}{\pi}\int d^2\beta \exp\left(-k\eta_s|\beta-\alpha|^2\right) P(\beta)\int d^2\beta \exp\left(-(1-k)\eta_s|\beta-\alpha|^2\right) P(\beta).
	\end{align}
	If we now recall the definition of the $s$-parametrized phase-space distributions in Eq.~\eqref{eq:sSupp}, we arrive at our final inequality conditions 
	\begin{align}
		P(\alpha;s)-\frac{\pi(1-s)}{2(1-k)k} P(\alpha;s_k) P(\alpha;s_{(1-k)})\stackrel{\mathrm{cl}}\geq 0
	\end{align}
	with $s_k=1-2/(k\eta_s)$ (or $s_k=1-(1-s)/k$), which is our main result -- connecting different $s$-parametrized  phase-space distributions to each other.
	Any violation of this inequality is a direct signature of the nonclassicality of the corresponding quantum state.

\section{Multimode generalization}
\label{sec:generalization}

    \begin{figure}[ht]
		\centering
		\includegraphics[width=0.5\columnwidth]{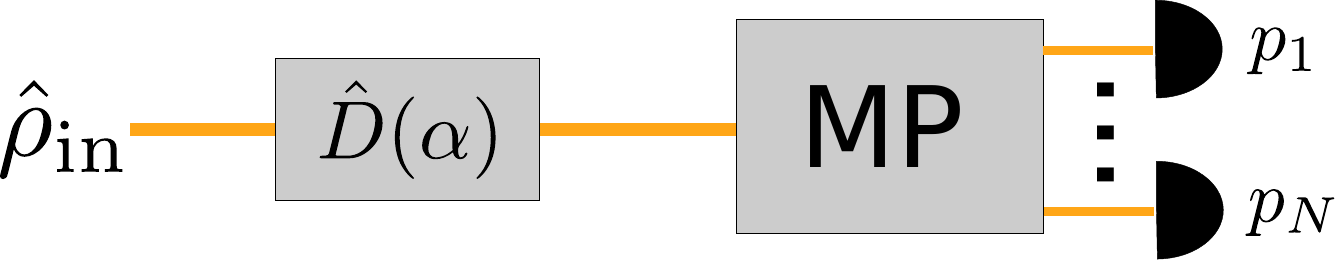}
			\caption{
			    General setup of multiplexed (MP) detection of the displaced quantum state.
			    This scheme generalizes the correlation measurement in the Fig. 3 a) of the main manuscript to $N$ detection channels.
			}
		\label{fig:multiplexing}
	\end{figure}
    
    Here, we generalize the relation between correlation measurements and the introduced phase-space-distribution inequalities to general multiplexing (MP) scenarios [cf. Fig \ref{fig:multiplexing}],  which is the setting of a usual multiplexed detection scheme \cite{paul_1996,kok_2001,achilles_2003,fitch_2003,castelletto_2007,schettini_2007,sperling_2015,bohmann_2018}.
	The multiplexing step transforms the input state [cf. Eq. (1) of the Letter] to an $N$-mode output state 
	\begin{align}
			\hat \rho_{\mathrm{out}}= \int d^2\beta P(\beta)|u_1\beta,\dots, u_N\beta\rangle\langle u_1\beta,\dots, u_N\beta|,
	\end{align}
	where the $u_i$ are the splitting rations with $\sum_{i=1}^N |u_i|^2=1$.
	The zero-count probability of the displaced state in  each channel is given by $p_i(\alpha,\eta |u_i|^2)=\langle{:}\exp[-\eta |u_i|^2 \hat n(\alpha)]{:}\rangle$ with $\eta$ being the detection efficiency of the detectors.
	We note that from each $p_i$ we can sample a corresponding $s$-parametrized phase-space distribution with $s_i=1-2/(\eta |u_i|^2)$ and the coincident zero-count detection of all channels, $p_{1,\dots,N}(\alpha,\eta)$, corresponds to the sampling of a distribution with $s=1-2/\eta$; cf. Eq.~(7) in the main manuscript.
	By applying Chebyshev's integral inequality \cite{mitrinovic_1970} $N{-}1$ times, we obtain the multimode zero-count condition 
	\begin{align}\label{eq:click}
	    p_{1,\dots,N}(\alpha,\eta)-\prod_{i=1}^N p_{i}(\alpha,\eta |u_i|^2 )\stackrel{\mathrm{cl}}\geq 0.
	\end{align}
	It is easy to see that this multimode consideration is not restricted to the consideration of correlation measurements only, but can be applied to any $s$-ordered phase-space distribution, which yields the multimode generalization of Eq.~(5) in the Letter
	\begin{align}
		P(\alpha;s)-\left(\frac{\pi}{\eta_s}\right)^{N-1}\prod_{i=1}^N k_i^{-1}  P(\alpha;s_{k_i}) \stackrel{\mathrm{cl}}\geq 0,
	\end{align}
	with $\sum_{i=1}^N k_i=1$ and $k_i=|u_i|^2$.
\end{widetext}	
\bibliography{biblio}
\end{document}